\documentclass[twocolumn,aps]{revtex4}
\usepackage{epsfig}
\newcommand{\beq}{\begin{eqnarray}}
\newcommand{\eeq}{\end{eqnarray}}
\begin{document}
\title{Role of Charged Gauge Fields in Generating Magnetic Seed Fields in Bubble Collisions during the Cosmological Electroweak Phase Transition}
\author{Trevor Stevens\\
Department of Physics, New Mexico State University, Las Cruces, NM 88003\\
Mikkel B. Johnson \\
Los Alamos National Laboratory, Los Alamos, NM 87545 \\
Leonard S. Kisslinger\\
Department of Physics, Carnegie-Mellon University, Pittsburgh, PA 15213\\
Ernest M. Henley\\
Department of Physics, University of Washington, Seattle, WA 98195\\
W.-Y Pauchy Hwang\\
Department of Physics, National Taiwan University, Taipei, Taiwan 106 R.O.C\\
Matthias Burkardt\\
Department of Physics, New Mexico State University, Las Cruces, NM 88003\\ }

\begin{abstract}
   We calculate the magnetic field generated during bubble collisions in a first-order electroweak phase transition that may occur for some choices of parameters in the minimal supersymmetric Standard Model. We show that for sufficiently gentle collisions, where the Higgs field is relatively unperturbed in the bubble overlap region, the equations of motion can be linearized so that in the absence of fermions the charged  $W^{\pm}$ fields are the source of the electromagnetic current for generating the seed fields. Solutions of the equations of motion for the charged gauge fields and Maxwell's equations for the magnetic field in O(1,2) space-time symmetry are expressed in closed form by applying boundary conditions at the time of collision.  Our results indicate that the magnetic fields generated by charged  $W^{\pm}$ fields in the collision are comparable to those found in previous work.  The magnetic fields so produced could seed galactic and extra-galactic magnetic fields observed today.

\end{abstract}
\maketitle
\noindent
PACS Indices:12.38.Lg,12.38.Mh,98.80.Cq,98.80.Jk
\vspace{1mm}


\vspace{1mm}
\noindent

\section{Introduction}
Identifying the source of the observed large-scale galactic and extra-galactic
magnetic fields remains an unresolved problem of astrophysics. One of the interesting possible sources is cosmological magnetogenesis, where the seed fields would have arisen during one of the early-universe phase transitions: the Quantum Chromodynamic chiral phase transition that led from the quark-gluon plasma to our present hadronic universe or the much earlier electroweak phase transition (EWPT) in which the Higgs and the other particles acquired their masses. See Ref.~\cite{gr} for a comprehensive review.

Here we consider magnetogenesis of the EWPT, which most likely requires a first-order phase transition.  Electroweak baroygenesis is a closely related problem of great interest and one requiring a first-order phase transition also. Although it is generally believed that there can be no first-order EWPT in the Standard Model~\cite{klrs}, there has been a great deal of activity in supersymmetric extensions~\cite{r90}, and for certain minimal extensions of the Standard Model there can be a first-order phase transition~\cite{laine,cline,losada}. One such extension that allows a first-order phase transition requires a right-handed Stop, the partner to the top quark, with a mass similar to the Higgs~\cite{laine,bodeker}. Other models for CP violation and baryogenesis have been proposed, including two-Higgs models~\cite{laine,cline,losada} and leptoquarks~\cite{herczeg}.

Limits on parameter space of the minimal extension of the Standard Model (MSSM) placed by electric dipole moment measurements and dark matter searches allow a first-order electroweak phase transition that could lead to successful electroweak baryogenesis~\cite{cpr} and the possibility that we are exploring, namely that magnetic seed fields responsible for the large-scale magnetic fields seen today are created during the era of the EWPT.  Accordingly, we consider the production mechanism of magnetic seed fields in a MSSM using an equation of motion (EOM) approach similar to one we developed in studying bubble collisions in the QCD phase transition~\cite{lsk2}.

The most detailed previous research on the study of magnetic fields during the EWPT along these lines was based on an  Abelian Higgs model~\cite{kv95,ae98,cst00}, in which the electromagnetic ({\it em}) field originated from gradients in the Higgs phase when bubbles of the broken phase collided and overlapped. In this paper, we calculate the magnetic field instead by starting from the MSSM using results of~\cite{jkhhs,hjk}, where we derive the EOM directly from the electroweak (EW) Lagrangian and express the {\it em}  current explicitly in terms of the charged $W^{\pm}$ fields appearing in the Lagrangian.

For cosmological magnetogenesis, the characteristics of the primordial magnetic field are vastly modified during cosmic evolution to the present day.  Much progress has been made in modeling this evolution in magnetohydronamics (MHD), first solving the equations for the period leading to the formation of galaxies and galactic clusters considering all relevant dissipative processes such as viscosity, and then for the evolution of these structures including the  possibility that they provide a large-scale dynamo. Such studies have led to quantitative predictions for magnetic field energy and coherence length at the present epoch. One detailed study of possible galactic or extra-galactic magnetic fields evolving from the EWPT showed that random seed fields would not
create such large-scale magnetic fields~\cite{son99}.  Results of more recent studies~\cite{jed}, however, are more optimistic and support the possibility that galactic cluster magnetic fields may in fact be entirely primordial in origin.
Our main interest in the present work is the mechanism that might have led to the creation of the seed fields during the EWPT rather than their cosmic evolution.

For our present exploratory work, the precise nature of the extension of the Standard EW Model is not needed as long as it supports a first-order phase transition. We begin, in Sect.~II, by reviewing the derivation of our EOM and explaining some of the considerations motivating our numerical work. In Sect.~III we review the previous work done in the  Abelian Higgs model. In our work as well as that of Ref.~\cite{kv95,ae98,cst00}, bubble nucleation and growth is driven by the dynamics of the Higgs field with an effective potential allowing a first-order phase transition, building on the classic work of Coleman~\cite{coleman}. Fermions are not explicitly considered. In Sect.~IV we make use of the observations of Sect.~II to simplify the theory along lines previously discussed in Ref.~\cite{jkhhs}, concentrating on a regime in which the bubble collisions may be considered "gentle". Restricting the application to gentle collisions has the advantage that the EOM linearize and display a more transparent connection to the work of \cite{kv95,ae98,cst00}. A closed-form solution of our EOM is given in Sect.~V, and the results of our calculation of the corresponding seed fields are given in Sect.~VI.

\section{Equation of Motion Approach to Magnetic Field Creation during the EWPT}

Our ultimate goal is to determine the magnetic seed fields by solving EOM derived from the MSSM, a goal shared with Ref.~\cite{hjk}. Both approaches obtain EOM derived from a similar Lagrangian, require a first-order phase transition, and take charged gauge fields as the source of the {\it em}  current, but beyond this there are important distinctions.  For example here we focus on a regime where linearized equations are appropriate, whereas in Ref.~\cite{hjk} the full nonlinear dependence on the $W$ fields is retained assuming an I-spin ansatz for the $W$ and for the {\it em}  field. Additionally, here magnetic fields generated in bubble collisions are obtained, whereas in Ref.~\cite{hjk} {\it em}  fields generated in {\it nucleation} are studied.  Reference~\cite{hjk} explicitly includes the right-handed Stop in the Lagrangian, whereas in the present work we integrate out all supersymmetric partners including the Stop.

\subsection{MSSM EW equations of motion}

In this section we summarize the derivation of our equations of motion~\cite{jkhhs}.  As stated above, we imagine our MSSM Lagrangian to support a first-order phase transition, but all supersymmetric partners of the Standard Model fields are integrated out, including the Stop.  This Lagrangian is
\beq
\label{L}
  {\cal L}^{MSSM} & = & {\cal L}^{1} + {\cal L}^{2}
\nonumber  \\
      && +{\rm leptonic \: and \: quark \: interactions }\\ \nonumber
         {\cal L}^{1} & = & -\frac{1}{4}W^i_{\mu\nu}W^{i\mu\nu}
  -\frac{1}{4} B_{\mu\nu}B^{\mu\nu} \\ \nonumber
 {\cal L}^{2} & = & |(i\partial_{\mu} -\frac{g}{2} \tau \cdot W_\mu
 - \frac{g'}{2}B_\mu)\Phi|^2  -V(\Phi) \nonumber \, ,
\eeq
with
\beq
\label{wmunu}
  W^i_{\mu\nu} & = & \partial_\mu W^i_\nu - \partial_\nu W^i_\mu
 - g \epsilon_{ijk} W^j_\mu W^k_\nu\\ \nonumber
 B_{\mu\nu} & = & \partial_\mu B_\nu -  \partial_\nu B_\mu \, ,
\eeq
where the $W^i$, with i = (1,2), are the $W^{\pm}$ fields, $\Phi$, is the Higgs field, $\tau^i$ is the SU(2) generator.  The electromagnetic and Z fields are defined as
\beq
\label{AZ}
   A^{em}_\mu &=& \frac{1}{\sqrt{g^2 +g^{'2}}}(g'W^3_\mu +g B_\mu) \nonumber \\
   Z_\mu &=& \frac{1}{\sqrt{g^2 +g^{'2}}}(g W^3_\mu -g' B_\mu) \; .
\eeq

The effective Higgs potential $V(\Phi )$ is not known at the present time,
depending as it does on unknown parameters of the MSSM as well as the properties of the plasma in the early Universe at the time of the EW phase transition. The various parameters are discussed in many publications~\cite{laine}.
Fortunately, the specific form of $V(\Phi )$ is not relevant for the purposes of this paper, beyond the requirement that it should produce a first-order phase transition as it would in an underlying MSSM extension including a light right-handed Stop~\cite{bodeker}. Later, in
an illustrative calculation, we use a simple form for $V(\phi)$ taken from Ref.~\cite{coleman}.
For our calculations we
need $g=e/\sin\theta_W = 0.646$, $g'=g \; \tan\theta_W =0.343$, and
$G=gg'/\sqrt{g^2 +g^{'2}}=0.303$.

In the picture we are using, the Higgs field plays a dynamic
role in EW bubble nucleation and collisions.
Our form for the Higgs field is
\beq
\label{phi}
         \Phi(x) & = & \left( \begin{array}{clcr} 0 \\
                                \phi(x)
         \end{array} \right) \; ,
\eeq
and thus
\beq
\label{tau}
  \tau \cdot W_\mu \Phi & = &
    \left( \begin{array}{clcr} (W^1_\mu-iW^2_\mu) \\
                            - W^3_\mu
         \end{array} \right) \phi(x)  \, .
\eeq
We also use the definitions
\beq
\label{phis}
       \phi(x) &\equiv& \rho(x)e^{i\Theta(x)} \nonumber \\
       |\phi(x)|^2 &=& \rho(x)^2 \nonumber \; .
\eeq
With these definitions $ {\cal L}^{2}$ is (j = (1,2,3))
\beq
\label{L2}
 {\cal L}^2 & = &  \partial_\mu \phi^*\partial^\mu \phi +
(i(\partial_\mu \phi^*)\phi - i\phi^* \partial_\mu \phi)
( -g W^3_\mu \nonumber \\
          && +g' B_\mu)/2 + \phi^* \phi[(\frac{g}{2})^2 (W^j)^2 +
(\frac{g'}{2})^2 B^2 \nonumber \\
 &&-\frac{gg'}{2} W^3 \cdot B]-V(\Phi) \; .
\eeq

The equations of motion are obtained by minimizing the action
\beq
\label{action}
 \delta \int d^4 x [{\cal L}^{1}+{\cal L}^{2} ] & = & 0 \, ,
\eeq
i.e., we do not include the leptonic parts of ${\cal L}$.
The result of doing this yields the following set of equations, where our metric is $(1,-1,-1,-1,-1)$.  The modulus $\rho$ of the Higgs field satisfies the ``$\rho$-equation",
\begin{eqnarray}
\label{eomf}
0 & = & \partial^2\rho (x)-\frac{ g^2}{4}\rho(x)[W^1\cdot W^1 \\ \nonumber
&& +W^2\cdot W^2]-\rho (x)\psi_\nu\psi^\nu+\rho (x)\frac{\partial V}
{\partial \rho^2}~,
\end{eqnarray}
where the quantity $\psi_\nu$ is defined in terms of the phase of the Higgs field, the $B$ field, and the $W^3$ field as
\begin{equation}
\label{eompsi}
\psi_\nu (x)\equiv \partial_\nu\Theta+\frac{g'}{2}B_\nu-\frac{g}{2}W^3_\nu~.
\end{equation}
It obeys a ``$\Theta$-equation",
\begin{equation}
\label{eomth}
0=\partial^\nu \rho (x)^2\psi_\nu (x),
\end{equation}
with $B$ field satisfying a ``$B$-equation",
\begin{equation}
\label{eomb}
0=\partial^2B_\nu-\partial_\nu \partial\cdot B+\rho (x)^2g'\psi_\nu (x).
\end{equation}
Finally, the $W$ field satisfies the two ``$W$-equations",
\begin{eqnarray}
\label{eomw3}
   & 0 & = \partial^2 W^3_\nu-\partial_\nu \partial\cdot W^3 -f(x)^2 g\psi_\nu(x)  \\ \nonumber
 & - &g \epsilon_{3jk}[ W^k_\nu \partial\cdot  W^j  + 2 W^j\cdot \partial  W^k_\nu
-W^j_\mu \partial_\nu  W^{k\mu}] \\ \nonumber
& + & g^2 \epsilon_{3jk} W^j_\mu \epsilon^{klm} W^{l\mu} W^m_\nu
\end{eqnarray}
for $i=3$, and
\begin{eqnarray}
\label{eomw12}
& 0 & = \partial^2 W^i_\nu-\partial_\nu \partial\cdot W^i
+\frac{1}{2}f(x)^2 g^2 W^i_\nu \\ \nonumber
 &-&g \epsilon_{ijk}[ W^k_\nu \partial\cdot  W^j
 + 2 W^j\cdot\partial  W^k_\nu
 - W^j_\mu \partial_\nu  W^{k\mu}] \\ \nonumber
& + & g^2 \epsilon_{ijk} W^j_\mu \epsilon^{klm} W^{l\mu} W^m_\nu
\end{eqnarray}
for $i=(1,2)$.

These are exact equations of motion in our MSSM model.

\subsection {First-order electroweak phase transition}

As we indicated, in this paper we study the magnetic field generated during a first-order electroweak phase transitions by finding solutions of the equations of motion derived above in the particular regime of gentle collisions.  Since the basis of our theory involves complicated coupled nonlinear partial differential equations (PDE), we would like to begin by illustrating some of the central ideas by discussing them within a model developed by Coleman~\cite{coleman} that is conceptually transparent but contains much of the complexity of the actual electroweak transition that we wish to examine.

In Coleman's model, the Higgs field is a real scalar field $\phi$ ($\Theta =0$) satisfying the nonlinear PDE
\begin{equation}
\label{fcoleman}
\partial^2\phi (x)+\phi(x)\frac{\partial V}
{\partial \phi^2}=0~.
\end{equation}
The effective potential $V(\phi )$, which specifies the dependence of the energy of vacuum on the scalar
field, has a metastable minimum for a value of $\phi =\phi_1$ separated from a second minimum at $\phi = \phi_2$ by a barrier. Coleman identifies the two minima with ``true" and ``false" states of the vacuum, respectively.  In his model, a symmetry breaking term is added to $V(\phi )$ to give the true vacuum a slightly lower energy. Initially, the system is imagined to exist in the false vacuum for which $\langle\phi\rangle=\phi_1$ but over time makes a transition to the true vacuum in which $\langle\phi\rangle=\phi_2$.

Coleman confirmed that in this model the system moves from the false to true vacuum by bubble nucleation as one expects of a first-order phase transition.  He showed that bubbles nucleate as tunneling (instanton) solutions of Eq.~(\ref{fcoleman}) connecting $\phi$ in the true and false vacua in Euclidean space-time, and he was able to calculate the nucleation probability.

During the actual electroweak phase transition in the early Universe, the Higgs couples to the other particles in the electroweak theory, and as a result these particles acquire mass in the true vacuum as the value of $\langle \phi \rangle$ changes from $\phi_1=0$ to $\phi_2 > 0$. Once nucleated, bubbles grow and collide as solutions of Eq.~(\ref{fcoleman}) in Minkowski space-time. Coleman did not consider the physics of collisions, but these collisions of central interest in our work because it is through them that the magnetic field will be generated.

\begin{figure}
\centerline{\epsfig{file=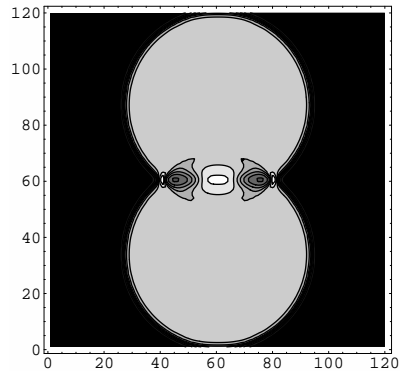, height=6.5cm,width=6.5cm}}
\caption{Two bubbles colliding in the Coleman model as explained in the text.  The distance scale is in arbitrary units.}
\label{Fig.1}
\end{figure}

\begin{figure}
\centerline{\epsfig{file=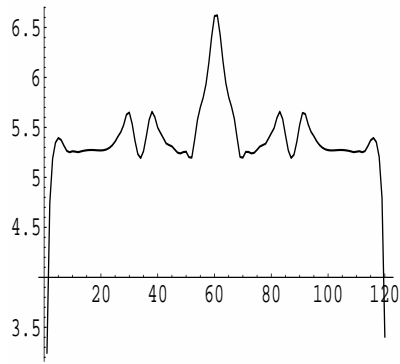,height=6.5cm,width=6.5cm}}
\caption{A vertical slice through the colliding bubbles shown in Fig. 1 as explained in the text.  The distance scale is in arbitrary units.}
\label{Fig.2}
\end{figure}

We have modeled bubble collisions by obtaining numerical solutions of Eq.~(\ref{fcoleman}) in 2 + 1 dimensions to illustrate an important feature of collisions.  We assume that the bubbles nucleate at rest at time $t=0$ with radii $r(t=0)=R_0$ small compared to their radii $r(t_c)=R_c$ at the time of collision. For $0<t<t_c$ the magnitude of the scalar field is constant at $\phi\equiv \rho_0=5$ (in arbitrary units) throughout each bubble, dropping rapidly to zero at its surface. Once the bubbles collide, the scalar field begins to fluctuate around the value $\rho_0$. The fluctuations are greatest for $r\approx R_c$, just after the surfaces first touch, corresponding to the situation shown in Fig.~\ref{Fig.1}.  A vertical slice through the symmetry axis is shown in Fig.~\ref{Fig.2}.  It is seen that at this point the scalar field fluctuates by as much as $30\%$ from $\rho=\rho_0$.  These fluctuations rapidly damp out and essentially disappear once the bubbles interpenetrate with $r(t)\approx R_0+R_c$.

For the purposes of this paper it is important that in the collision these fluctuations in the scalar field remain small compared to $\rho_0$. We refer to a collision of this character as a gentle collisions.  We will make use of the observation that collisions between bubbles tend to be gentle to guide our approach to show how magnetic field generation arises as a result of charged  $W^{\pm}$ dynamics as the bubbles collide.

To model the dynamics of a first-order electroweak phase transition of the early Universe, one needs to consider the dependence of the effective potential $V(\Phi ,T)$ on temperature $T$ (see Ref.~\cite{eikr} for a comprehensive phenomenological study of the kinetics of cosmological first-order phase transitions such as the EW phase transition in terms of such an effective potential).  Just as imagined in Coleman's toy model, the vacuum state of the Universe corresponds to a local minimum in $V(\Phi ,T)$.  For T greater than a critical temperature $T_c$, the minimum in the effective potential is at $\langle\Phi\rangle=0$.  At $T_c$, $V(\Phi ,T)$ develops a degenerate second minimum at a value $\langle \Phi\rangle >0$ separated from the firsr by a barrier. As the Universe continues to expand and cool, the second minimum becomes deeper, meaning that the Universe begins to tunnel from the original, now metastable, false vacuum to the true vacuum by bubble nucleation. In Ref.~\cite{eikr} $T_c$ is estimated to be $T_c\approx 166$~GeV.  Because the two minima are separated by a barrier, the transition is delayed, a process referred to as supercooling.  As the phase transition develops, the bubbles expand with a speed $v$ and eventually completely merge, at which point the phase transition is completed. The terminal speed depends on the parameters of the phase transition but appears to be quite low, perhaps $v_{wall}\approx .1c$, as discussed in Ref.~\cite{dyne}.

Our theory is more complicated than that of Coleman since it based on the MSSM, with the Higgs coupled to the other fields of that theory.
We note here, by reference to Eq.~(\ref{eomf}), that the gentle character of the collision illustrated in Fig.~\ref{Fig.2} may change when the $W$ fields become large, with the Higgs field showing strong variation in the collision region.  We refer to collisions of this character as violent collisions, the subject of a future publication.

\section{Bubble Collisions in the Abelian Higgs Model}

The  Abelian Higgs model has been of interest as a prototype for the generation of magnetic fields in
the early universe in collisions of bubbles during a first-order EW phase
transition~\cite{kv95,ae98,cst00}.  The Lagrangian of the  Abelian Higgs model describes a complex scalar field coupled to the {\it em}  field $ A^{em}_\mu$.  It corresponds to the Lagrangian Eq.~(\ref{L}) in the  abelian sector, formed by eliminating the $W$ fields,
identifying $ A^{em}_\mu$ with the field $B_\mu$, and relating the electric charge $e$ to coupling
parameter $g'$ as $e=g'/2$.  In this limit, the equations of motion for the Higgs and magnetic field may be read off the results in
Eqs.~(\ref{eomth}, \ref{eomb},\ref{eomf}), {\it i.e.},
\begin{equation}
\partial^\nu \rho (x)^2\psi_\nu (x)=0~,
\end{equation}
\begin{equation}
\partial^2B_\nu-\partial^\mu\partial_\nu B_\mu+2e\rho (x)^2\psi_\nu (x) =0,
\end{equation}
and
\begin{equation}
\partial^2\rho (x)-\rho (x)\psi_\nu\psi^\nu+\rho (x)\frac{\partial V}
{\partial \rho^2}=0~.
\end{equation}
The quantity $\psi_\nu$ is now
\begin{equation}
\psi_\nu (x)= \partial_\nu\Theta+eA^{em}_\nu~.
\end{equation}
In this case, the phase transition is driven by the dynamics of the Higgs field along the lines studied by
Coleman~\cite{coleman} and discussed in connection with Fig.~\ref{Fig.1} and Fig.~\ref{Fig.2} in the previous section.

\subsection{ Gentle bubble collisions in the  Abelian Higgs model }
In their analysis of the  Abelian Higgs model, Kibble and Vilenkin~\cite{kv95} suggested one way in which magnetic fields
might be generated as bubbles collide. They considered the regime of gentle collisions, where $\rho(x)$ remains
constant, or nearly constant, in the region of overlap of the colliding bubbles.  They were able to gain insight into the
generation of magnetic fields in this case by making an expansion about point $\rho(x)=\rho_0$, which we shall refer to as
the Kibble-Vilenkin point.  For the case of gentle collisions, we can assume the following expansion,
\begin{eqnarray}
\label{kv1}
\rho (x)&=&\rho_0+a\delta \rho(x)\\
\psi_\nu(x)&=&a\psi^{(1)}+a^2\psi^{(2)}_\nu +~...~, \nonumber
\end{eqnarray}
where $a$ is the magnitude of $\rho_0-\rho (x)$ and is small by assumption (refer to the discussion in connection with Fig.~\ref{Fig.1} and Fig.~\ref{Fig.2}).  Substituting the expansion into the equations
of motion and requiring that the equations be satisfied at each order in the expansion parameter $a$, the relevant
$\Theta$ and $B$ equations give, to leading order in $a$, the following results
\begin{eqnarray}
\label{kv2}
(\partial^2+2e^2\rho_0^2)\psi^{(1)}_\nu=0\\
\label{kv3}
\partial^\nu\psi^{(1)}_\nu=0~,
\end{eqnarray}
where now
\begin{eqnarray}
\label{kv4}
\psi^{(1)}_\nu=\partial_\nu\Theta+eA^{em(1)}_\nu~.
\end{eqnarray}
From these equations one can easily derive the equations of Kibble and Vilenkin~\cite{kv95} in the axial gauge for $A^{em}_\mu$.

\subsection{Bubble collisions and $O(1,2)$ symmetry}
Prior to nucleation, the dynamics is most easily formulated in the Euclidean metric, which is $O(4)$ symmetric.  After nucleation, the bubble expands in Minkowski space-time, which is $O(1,3)$ symmetric~\cite{coleman}.  For collisions, Kibble and Vilenkin study the case of two bubbles nucleated simultaneously at time $t=0$ at positions symmetrically located about the origin on the $z-$ axis.  In the notation of Ref.~\cite{cst00}, the radii of the bubble at nucleation are $R_0$.  Once nucleated, the bubbles expand with radii $r(t)$ according to
\beq
\label{bubradius}
r(t)^2=R_0^2+t^2~.
\eeq
The bubbles collide on the z axis at $t=t_c$ with radii $r(t_c)=R_c$, and hence the nucleation points are at $z=\pm R_c$. In this system, the collision of two bubbles has the symmetry $O(1,2)$ and may be expressed in the coordinates $(\tau,z )$, where
\beq
\label{geom}
\tau^2=t^2-x^2-y^2~.
\eeq

Kibble and Vilenkin formulated their  Abelian Higgs model in terms of the variables $(\tau,z )$ and found the $O(1,2)$ symmetric solutions of Eqs.~(\ref{kv2}) and (\ref{kv4}).  They demonstrated from them that when the phase of the Higgs fields is initially different within each
bubble an axial magnetic field forms as the bubbles collide and that this field has the structure of an expanding ring encircling the overlap region of the colliding bubbles.

\subsection{Finite wall speed and plasma conductivity}
The  Abelian Higgs model has the feature that the bubbles accelerate freely to the speed of light along space-time hyperboloids. Corrections come from a variety of sources and were discussed comprehensively in Refs.~\cite{kv95,ae98,cst00}, and the effects of these corrections on the subsequent evolution of the magnetic field was estimated. We discuss these results briefly next.

One of the important corrections is that bubble walls reach a terminal speed $v_{wall}<<1$ as the bubbles experience collisions with constituents of the plasma. Another important correction is the conductivity of the medium.  The effects of finite conductivity lead to decay of the currents (and therefore the magnetic field) with a characteristic time $t_d\approx \sigma/m^2$~\cite{kv95}.  However, with values of conductivity that are believed to characterize the plasma, currents and magnetic fields persist on time scales that are long compared to those of the symmetry breaking scale.

Another consequence of the large conductivity arises from the following considerations.  Since the magnetic fields propagate with the speed of light, in the absence of conductivity, for slowly expanding bubbles these fields would very quickly escape from the region of the bubble collision and move into the surrounding false vacuum.  However, because of the large conductivity the magnetic fields become "frozen" or confined to the interior of the bubbles.  This is particularly important in the case of slowly moving bubble walls, where this effect prevents the escape of magnetic flux into the false vacuum.  Kibble and Vilenkin showed that the loss of flux is negligible provided that $\sigma R_c v >>1$, where $R_c$ is the bubble radius at collision time.

\section{Magnetic Field Generation in the MSSM in Gentle Collisions }

The main difference of our current work compared to the  Abelian Higgs model arises from the fact that in our MSSM model the source of the current is the charged gauge fields. These  $W^{\pm}$ originate in the plasma and populate the bubble by passing through the bubble wall following nucleation.  As they do this, they acquire mass and thereby loose kinetic energy.  We envision them to cool as the bubble expands, thereby occupying a low-energy mode, the occupation of which grows as the bubble expands. We assume in this work that early in the phase transition all such  $W^{\pm}$ entering the bubble drop into this mode, which has a high degree of coherence, much like a state of electrons in a superconductor (except that the $W$ are bosons). It is quite different from the more familiar thermal modes, which are incoherent.

As we shall see, this mode plays a special role in the theory. A significant consequence of the strong coupling to the Higgs field is that the $W$ fields follow the evolution of the Higgs field, and this is described by our EOM. We solve our EOM the magnetic field in gentle bubble collisions for the MSSM in parallel to the  Abelian Higgs model by applying simple boundary conditions in Kibble-Vilenkin geometry at the time of collision.
From our results we will be able to ascertain the relative importance of the charged  $W^{\pm}$ in the MSSM as compared to the Higgs mechanism of the  Abelian Higgs model as a source of magnetic field generation.

The dynamics of the coupled fields is clearly a non-equilibrium component of the phase transition, and eventually the coherence must dissipate as the system thermalizes. Determining the value time scale of the
dissipation is an important issue that we do not address here. We assume that the coherent evolution of these fields after the collision give rise to a magnetic field before thermal equilibrium is re-established.

It is our eventual goal to determine the magnetic field collisions between individual bubbles possessing O(3) symmetry as they would when evolving from nucleation with a finite wall velocity. Thus, in the
future we will examine the more numerically intensive case of collisions by solving the equations of motion numerically. Numerical methods for dealing this situation will be reported in forthcoming publications~\cite{stevens1}.

Here, we first derive, in Sect.~IV.A, the EOM in the case of gentle collisions. In the MSSM, the EOM are complicated nonlinear PDE coupling the $W$, $B$,
and $\phi$ (Higgs) fields. From the solution of these equations, the physical $Z$ and $A^{em}$ fields are determined by Eq.~(\ref{AZ}). As shown in Sect.~IV.B, this leads in turn to an expression for the electromagnetic current in terms of the  $W^{\pm}$ and to a corresponding Maxwell equation.

\subsection{EOM for gentle collisions in electroweak theory }
As indicated, we consider here only the regime of
gentle collisions in the context of the full MSSM.  This leads to a linearized theory giving EOM for these fields. Although the linearization leads to the same limitations as the  Abelian Higgs model discussed at the end of the last section, fortunately the corrections for the finite wall propagation and the finite conductivity of the plasma are easily estimated since the considerations  are similar to those of the  Abelian Higgs model.

As in the  Abelian Higgs model, the MSSM equations simplify upon expansion about the Kibble-Vilenkin point,
\begin{eqnarray}
\rho(x)=\rho_0+a\delta \rho (x).
\end{eqnarray}
The fact that $\psi$ and $W^d$ (for $d=(1,2)$) enter quadratically in the $\rho$-equation places two important constraints on
these quantities:  (1) $\psi$ and $W^d$ must have an expansion in odd powers of $a^{1/2}$, if we
require the square of these quantities be analytic in $a$; and, (2) expanding this equation to leading order in $a^{1/2}$,
we find that the terms $\psi^{(0)}$, $w^{(0)1}$, and $w^{(0)2}$ must vanish.  This is most easily seen in the Euclidean
metric, from the fact that the square of each enters with the same sign.  However, the same must be true in the Minkowski
metric as well by analytic continuation.  In view of these considerations, $\psi_\nu$ and
$W^d_\nu$ for $d=(1,2)$ have the following expansion
\begin{eqnarray}
\label{expand1}
\psi_\nu(x)=a^{1/2}\psi^{(1)}_\nu+a^{3/2}\psi^{(3)}_\nu +~...
\end{eqnarray}
\begin{eqnarray}
\label{expand2}
W^d_\nu=a^{1/2}w^{(1)d}_\nu+a^{3/2}w^{(3)d}_\nu +~...~.
\end{eqnarray}
It is natural that an expansion in the same parameter $a^{1/2}$ remains appropriate for $d=3$.  However,
there is no requirement that the leading term vanish, so we take
\begin{eqnarray}
\label{W3}
W^3_\nu =w^{(0)3}_\nu+a^{1/2}w^{(1)3}_\nu+a^{3/2}w^{(3)3}_\nu +~...~.
\end{eqnarray}

The $B$-, $\Theta$-, and $W$-equations then give, to first order in $a^{1/2}$,
\begin{eqnarray}
\label{eompsi1}
[\partial^2+\frac{\rho_0^2}{2}(g^2+g'^2)]\psi^{(1)}_\alpha=0\\
\label{eompsi2}
\partial^\alpha\psi^{(1)}_\alpha=0
\end{eqnarray}
where now
\begin{eqnarray}
\label{psi3}
\psi^{(1)}_\alpha(x)=\partial_\alpha\Theta-\frac{(g^2+g'^2)^{1/2}}{2}Z^{(1)}_\alpha~.
\end{eqnarray}
Comparing these equations to those in the  Abelian Higgs model, Eqs.(\ref{kv2},\ref{kv3},\ref{kv4}), we see that the
$Z$ field here plays the same role as the {\it em}  field did in that case.  Specifically, in this case, the phase
difference of the Higgs field now determines the gauge field $Z_\mu$ of Eq.(\ref{AZ}) within the bubble
overlap region.  Thus, for gentle collisions,  the mathematical problem in leading order
is no more complicated that it was in the  abelian case.

Equations for $w^{(1)d}_\nu$ may be obtained by expanding the $B$- and $W$-equations through order $a^{1/2}$.
For $d=$ 1 or 2 (corresponding to $d^{\prime}=$ 2 or 1, respectively), we obtain
the pair of equations
\begin{eqnarray}
\label{eomu12}
0&=&\partial^2w^{(1)d}_\nu-\partial_\nu\partial\cdot w^{(1)d}+ m^2w^{(1)d}_\nu \\ \nonumber
&-&2[\partial^\mu(w^{(0)3}_\nu w^{(1)d^{\prime}}_\mu -w^{(1)d^{\prime}}_\nu w^{(0)3}_\mu)\\ \nonumber
&+&(w^{(1)d^{\prime}}_\mu\partial^\mu w^{(0)3}_\nu -w^{(0)3}_\mu\partial^\mu w^{(1)d^{\prime}}_\nu) \\ \nonumber
& - &(w^{(1)d^{\prime}}_\mu\partial_\nu w^{(0)\mu 3} -w^{(0)3}_\mu\partial_\nu w^{(1)\mu d^{\prime}})]\\ \nonumber
&-&4[(w^{(0)3})^2w^{(1)d}_\nu-w^{(0)3}\cdot w^{(1)d}w^{(0)3}_\nu] ~,
\end{eqnarray}
where $m$ is the mass of the $W$ field and is given by
\beq
\label{wmass}
m^2=\frac{\rho_0^2g^2}{2}~.
\eeq
The corresponding equation determining $w^{(1)d}_\nu$ for $d=3$ is
\begin{eqnarray}
\label{eomu31}
\partial^2w^{(1)3}_\nu-\partial_\nu\partial\cdot w^{(1)3}=\rho_0^2g\psi^{(1)}_\nu~,
\label{defu}
\end{eqnarray}
which can be solved once the driving term
$\psi^{(1)}(x)$ has been independently determined from the solution of
Eqs.~(\ref{eompsi1},\ref{eompsi2}).

Note that knowledge of the field $w^{(0)d}_\nu$ for $d=3$ is required in order to solve Eq.~(\ref{eomu12}). This field is found to be the solution of
\begin{eqnarray}
\label{eomu30}
\partial^2w^{(0)3}_\nu-\partial_\nu\partial\cdot w^{(0)3}=0~.
\end{eqnarray}
Because the mass term is absent in this equation, clearly $w^{(0)3}$ can not contribute to Eq.~(\ref{W3}) inside the bubble where $m$ is non-zero. Thus $w^{(0)3}$ must vanish inside the bubble. Consequently, the equations for $w^{(1)1}$ and $w^{(1)2}$ simplify and become
\beq
\label{eomus12}
0=\partial^2w^a_\nu-\partial_\nu\partial\cdot w^a
+m^2w^a_\nu~,
\eeq
and we see that for sufficiently gentle collisions, {\it all} relevant equations are linear inside the bubble.

\subsection{The electromagnetic current for gentle collisions}
We may find the Maxwell equation for the electromagnetic field $A^{em}_\nu(x)$ by taking the linear combination of the $W^{(3)i}$ and and $B$ indicated in Eq~(\ref{AZ}).  Using Eqs.~( \ref{eomb}),(\ref{eomw3} ), an expression for the corresponding {\it em}  current $j_\nu(x)$ immediately follows.  It consists of terms quadratic and cubic in the three fields $W^i(x)$.

This result for $j_\nu(x)$ may again be simplified by expanding the $A^{em}$ and $W$ fields in powers of $a^{1/2}$.  Letting $a^{(n)}_\nu (x)$ refer to the terms in the expansion of $A^{em}_\nu(x)$, we find that the leading term of $A^{em}_\nu(x)$ is $a^{(2)}_\nu(x)$, satisfying the following Maxwell Equation,
\begin{eqnarray}
\label{maxwell}
\partial^2a^{(2)}_\nu &-&\partial_\nu\partial\cdot a^{(2)} \\ \nonumber
&=&\frac{gg'\epsilon^{ab3}}{(g^2+g'^2)^{1/2}}(w_\nu^{(1)b}\partial\cdot w^{(1)a} \\ \nonumber
&-&w^{(1)a}_\mu\partial_\nu w^{(1)\mu b}+2w^{(1)a} \cdot \partial w^{(1)b}_\nu ) \\ \nonumber &\equiv & 4\pi j_\nu (x)
~,
\end{eqnarray}
From this we learn that the first non-vanishing contribution to the {\it em}  current is of order $a^{3/2}$ and that it depends on the components $w^{(1)i}_\nu$ of the {\it charged} $W^i$ fields ($i=$ 1 and 2), calculated at order $a^{1/2}$.  Expressing the current in terms of $w^i$ of Eq.~(\ref{defu}), we find

\begin{eqnarray}
\label{current}
4\pi j_\nu (x)&=&
\frac{gg'\epsilon^{ab3}}{(g^2+g'^2)^{1/2}}(w^{(1)b}_\nu \partial\cdot w^{(1)a} \\ \nonumber
&-&w^{(1)a}_\mu\partial_\nu w^{(1)\mu b}+2w^{(1)a}\cdot \partial w^{(1)b}_\nu)
~,
\end{eqnarray}
It is easy to prove that this current is conserved,
\beq
\label{consj}
  \partial\cdot j(x)=0~,
\eeq
since $w^a_\nu (x)$ and $w^b_\nu (x)$ that appear in Eq.~(\ref{current}) commute with each other and satisfy the equation of motion given in Eq.~(\ref{eomus12}).

One of the most important features of the derivation is that the source current  of the electromagnetic field is given by the charged gauge $W^{\pm}$ fields. This is expected physically,
and is in sharp contrast with the Kibble-Vilenkin~\cite{kv95},
Ahonen-Enqvist~\cite{ae98}, Copeland-Saffin-T$\ddot{o}$rnkvist~\cite{cst00}
picture, in which the source for the
electromagnetic field arises entirely from the Higgs.

It should be clear from the fact that the electromagnetic current in Eq.~(\ref{current}) is antisymmetric in the labels $a$ and $b$ that this current will vanish unless the field $W^a(x)$ for $a=1$ has a different dependence on $x$ from that for $a=2$.  Thus, in an isolated bubble the electromagnetic current will vanish if $W^1(x)$ and $W^2(x)$ in this bubble differ at most by a phase.  However, when two bubbles meeting this condition collide, it is in general not the case that $W^a(x)$ satisfying the equations of motion will be proportional in the region of overlap.
This is the reason why magnetic fields are in general produced when bubbles collide.  We model this below by solving the equations for $W^a$ with different sets of boundary conditions on $W^a$ for $a=1$ and $a=2$.

\section{Solution of the MSSM Equations of Motion for Bubble Collisions in $O(1,2)$ Symmetry}
Following Kibble and Vilenkin, in Sect.~V.A we express the EOM using the $(\tau,z )$ variables and find the general solution of the EOM for $W^{\pm}$ fields for a pair of colliding bubbles.  In Sect.~V.B we find the solutions for specific boundary conditions.  In Sect.~V.C we determine the current corresponding to these solutions and find expressions for the corresponding magnetic field by solving Maxwell's equations.

\subsection{W equations with O(1,2) symmetry}
To express the equations of motion, Eq.~(\ref{eomus12}), in terms of the $(\tau,z)$ coordinates we define
\beq
\label{wdef}
w^{ a}_\nu (x)&=&w^a_z(\tau ,z),~~\nu=3 \nonumber \\
w^{ a}_\nu (x)&=&x_\nu w^a(\tau ,z),~~\nu=(0,1,2)~,
\eeq
with $w^a_z=-w^{za}$.  (In the remainder of the paper, we will find it convenient to use $\alpha$ to denote the Lorentz index for the values $\nu=(0,1,2)$.) We then find
\beq
\label{eqwz}
(\frac{\partial^2}{\partial \tau^2} +\frac{2}{\tau}\frac{\partial}{\partial \tau} -\frac{\partial^2}{\partial z^2} +m^2)w^{za}(x)=0
\eeq
and
\beq
\label{eqw}
(\frac{\partial^2}{\partial \tau^2} +\frac{4}{\tau}\frac{\partial}{\partial \tau} -\frac{\partial^2}{\partial z^2} +m^2)w^{a}(x)=0~,
\eeq
where we have used the fact that
\beq
\label{aux}
\partial\cdot w^a(x)\equiv \frac{\partial w^{za}}{\partial z}+3w^a+\tau \frac{\partial w^a}{\partial \tau}=0~.
\eeq
Equation~(\ref{aux}) is a consequence of Eq.~(\ref{eomus12}) and follows by taking the four-divergence of this equation.  Comparing our results to the  Abelian Higgs model, we see that $w^a$ takes over the role of the electromagnetic vector potential and $\partial_z w^{za}$ takes over the role of $-m/e\Theta$.  In our theory, to find the electromagnetic field we have the
additional steps of constructing the electromagnetic current in Eq.~(\ref{current}) and then solving Maxwell's equation, Eq.~(\ref{maxwell}).

The solution of Eqs.~(\ref{eqwz},\ref{eqw}, and \ref{aux}) is found by standard methods.  Expressing $W^a_\nu(x)$ as a Fourier transform in $z$, Eqs.~(\ref{eqwz},\ref{eqw}) give ordinary differential equations for the $\tau$-dependence, yielding
\begin{eqnarray}
\label{wsol}
w^a(\tau ,z)&=&\frac{1}{\tau^2}\sqrt{\frac{2}{\pi}}\int^\infty_{-\infty}\frac{e^{ikz}}{\sqrt{\omega_k}}( \\\ \nonumber
&a_k&(\cos \omega_k\tau - \frac{\sin \omega_k\tau }{\omega_k\tau }) - \\ \nonumber
 &b_k&(\sin \omega_k\tau + \frac{\cos \omega_k\tau }{\omega_k\tau }))dk
\end{eqnarray}
and
\begin{eqnarray}
\label{wzsol}
w^{za}(\tau ,z)&=&\frac{1}{\tau}\sqrt{\frac{2}{\pi}}\int^\infty_{-\infty}\frac{e^{ikz}}{\sqrt{\omega_k}}( c_k\sin \omega_k\tau \\ \nonumber
&+& d_k\cos \omega_k\tau )dk
\end{eqnarray}
We may use the auxiliary condition of Eq.~(\ref{aux}) to relate the coefficients of the linearly independent functions in the expansion of $w^a$ and $w^{za}$. This gives
\beq
\label{coefc}
ic_k=a_k\omega_k/k~,
\eeq
and
\beq
\label{coefd}
id_k=b_k\omega_k/k~.
\eeq

\subsection{Boundary conditions}
As we indicated earlier, the relative phase of the two charged $W^\pm$ fields is irrelevant for a single bubble, but in a collision, this is not so. To be specific, let us assume that when the colliding bubbles first touch at $z=0$, the magnitude of the $W^a_z$ fields is the same in each bubble. It will sometimes happen, though, that the sign of $W^{+}_z$ is opposite in the two colliding bubbles while $W^{-}_z$ the same sign. We will refer to the first situation, where $W_z$ changes sign across the point of contact at $z=0$, with as superscript I, and the second situation with a superscript II.

The two situations correspond to distinct solutions of the EOM, each described by its own boundary condition. The boundary condition BCI for the first situation is
\begin{eqnarray}
\label{BCI}
w^{zI}(\tau =t_c,z)&=&w~\epsilon (z)\\ \nonumber
\frac{\partial }{\partial\tau} w^{zI}(\tau =t_c,z)&=&0~.
\end{eqnarray}
where $\epsilon (z)$ is the sign of $z$, and that for the second situation BCII,
\begin{eqnarray}
\label{BCII}
w^{zII}(\tau =t_c,z)&=&w \\ \nonumber
\frac{\partial }{\partial\tau} w^{zII}(\tau =t_c,z) &=&0~.
\end{eqnarray}
We note in passing that, from the definition of $\tau$ in Eq.~(\ref{geom}), the boundary conditions in Eqs.~(\ref{BCI}) and (\ref{BCII}) specify the value of $w^z$ on a cylinder of radius $\sqrt{x^2+y^2}=\sqrt{t^2-t_c^2}\equiv b(t)$, where $b(t)$ is the radius of the circle of intersection of the colliding bubbles at time $t$.

The constant $w$ in Eqs.~(\ref{BCI} and \ref{BCII}) fixes the density of the charged gauge fields  $W^{\pm}$ inside the bubble; charge neutrality requires that $w$ is the same for both $a=1$ and $a=2$.  We determine $w$ in the appendix in a model that assumes the average number density of the  $W^{\pm}$ quanta inside the bubble (in the true vacuum) is constant in time and equal to the number density of those  $W^{\pm}$ quanta in the thermal plasma that can make a transition into the bubble without violating overall energy-momentum conservation. Since the charged gauge field inside the bubble is normalized to the one outside the bubble, we should identify $W^d_\nu$ in Eq.~(\ref{expand2}) with $w^{(1)d}_\nu$ rather than with $a^{1/2}w^{(1)d}_\nu$.

From Eqs.~(\ref{wsol},\ref{wzsol}) we can determine the full $(\tau ,z)$ dependence of the fields $w^{za}$ and $w^a$ corresponding to each of the two boundary conditions. We will then calculate the magnetic field from a collision of the two bubbles as an indication of what magnetic fields can be expected when they arise from the current of the charged  $W^{\pm}$ fields.

Equating Eq.~(\ref{wzsol}) and its $\tau$ derivative at $\tau=t_c$ provides, in a straightforward fashion, the coefficients $a_k$ and $b_k$ for BCI and BCII.  The solution is most direct for BCII, leading to
\begin{eqnarray}
\label{sBCII}
w^{zII}(\tau,z)&=&\frac{w}{m\tau}(\sin mT+mt_c\cos mT) \\ \nonumber
w^{II}(\tau,z) &=&0~,
\end{eqnarray}
where $T=\tau -t_c$.
For BCI, we may perform the sum over the modes $k$ to obtain
\begin{eqnarray}
\label{sBCI}
w^{zI}(\tau,z)&=&\frac{wt_c}{\tau}\epsilon (z)\Theta(T-|z|)\int^{|z|}_0 \\ \nonumber
&(&\frac{1}{t_c}+\frac{\partial}{\partial T})J_0(m\sqrt {T^2-z'^2}dz'\\ \nonumber
&+&\Theta(|z|-T)\frac{1}{m}(\frac{1}{t_c}+\frac{\partial}{\partial T}) \sin mT \\ \nonumber
w^{I}(\tau,z) &=&-\frac{wt_c}{\tau^2}\Theta (T-|z|)(J_0(m\sqrt {T^2-z^2}) \\ \nonumber
&+&\frac{\sqrt {T^2-z^2}}{mt_c\tau}J_1(m\sqrt {T^2-z^2}))~.
\end{eqnarray}

\subsection{Magnetic field}
To obtain the magnetic field we need to solve Maxwell's equation, Eq.~(\ref{maxwell}), with the current given by Eq.(\ref{current}), and with the  $W^{\pm}$ fields appearing in the current given by the solutions of Eqs.~(\ref{eqwz}, \ref{eqw}, and \ref{aux}).  The solution of Maxwell's equation is completely determined once we specify the boundary conditions on the $A^{em}$ field with the  $W^{\pm}$ fields given above in Eqs.~(\ref{sBCII}, \ref{sBCI}) .

In the $(\tau,z)$ coordinates, the current in Eq.~(\ref{current}) may be written in the form
\beq
\label{formcurrent}
j_\nu(x)=(j_z(\tau,z),x_\alpha j(\tau,z))~.
\eeq
Evaluating the current with  $W^{\pm}$ fields satisfying the boundary conditions given in Eqs.~(\ref{BCII} and \ref{BCI}) and using the auxiliary condition of Eq.~(\ref{aux}),  we find that Eq.~(\ref{current}) simplifies and becomes
\begin{eqnarray}
\label{jfinal}
4\pi j(\tau ,z)&=& \frac{gg'}{\sqrt{g^2+g'^2}} \frac{1}{\tau}(-2\tau w^{zII}\frac{\partial w^{I}}{\partial z} \\ \nonumber
&+& w^{zI}\frac{\partial w^{zII}}{\partial \tau }- w^{zII}\frac{\partial w^{zI}}{\partial \tau })
\end{eqnarray}
and
\begin{eqnarray}
\label{jzfinal}
4\pi j_z(\tau ,z)&=&-\frac{gg'}{\sqrt{g^2+g'^2}}( 2\tau w^{I}\frac{\partial w^{zII}}{\partial \tau} \\ \nonumber
&-& w^{zII}\frac{\partial w^{zI}}{\partial z }) ~.
\end{eqnarray}
When evaluating the partial derivatives in Eq.~(\ref{jfinal},\ref{jzfinal}) using the expressions in Eqs.~(\ref{sBCII}, \ref{sBCI}) we do not let the derivatives act on the $\theta$ functions. We note that ignoring the surface derivatives is quite consistent with current conservation and should therefore lead to a valid expression for the magnetic field throughout the bubble interior.

Because the electromagnetic current has the form given in Eq.~(\ref{formcurrent}), the electromagnetic field has this form also,
\beq
\label{formemfield}
a_\nu(\tau,z )=(a_z(\tau,z),x_\alpha a(\tau,z))~.
\eeq
Maxwell's equation becomes quite simple in the $(\tau,z)$ with the axial gauge, $a_z=0$, namely,
\beq
\label{afield}
-\frac{\partial^2 }{\partial z^2}a(\tau,z)=4\pi j(x,\tau)~.
\eeq

Applying the boundary conditions on the $a(\tau,z)$ field, namely $a(t_c,z)=0$ and $\partial_z a(\tau=0,z)$, we find
\beq
\label{afield}
a(\tau,z)=-4\pi\int_{-\infty}^zdz'\int_{-\infty}^{z'}j(\tau,z'')dz''~.
\eeq
The magnetic field $B^i$ is the curl of $A^{\i~em}$ where $A^{\i~em}=(x a(\tau,z),y a(\tau,z),0)$ in the axial gauge with $a(\tau,z)$ given in Eq.~(\ref{afield}).  Thus,
\begin{eqnarray}
\label{Bfield}
B^z&=&0,\\ \nonumber
B^x&=&-4\pi y \int_{-\infty}^{z}j(\tau,z')dz'~.\\ \nonumber
B^y&=&4\pi x \int_{-\infty}^{z}j(\tau,z')dz'~.\\ \nonumber
\end{eqnarray}
Clearly, as in the work of Kibble and Vilenkin, the $B$ field is entirely azimuthal, encircling the axis of collision.

\section {Numerical Results}
In this section we compute the magnetic field using the theory developed above, which assumes a non-conducting medium $\sigma=0$ and a terminal wall speed $v_{wall}=1$.  We will compare our magnetic fields to the results obtained in the  Abelian Higgs model under the same assumptions.

Although, as we admitted earlier, the assumptions of $\sigma=0$ and $v_{wall}=1$ are unrealistic for the actual EWPT, the corrections have been thoroughly studied in Refs.~\cite{kv95} and \cite{ae98}.  The corrections are therefore well-understood, and they will be easy to apply to our results since they are not specific to the source of the electromagnetic current. We will also do this below.

We first show our azimuthal field $B^\phi$ using our results in Eqs.~(\ref{BCI}, \ref{BCII},\ref{jfinal}, and \ref{Bfield}  ). The mass appearing in the expressions for the  $W^{\pm}$ fields is of course the mass of the $W$ boson, $m_W= 80.6~GeV$.  To facilitate the comparison to the Abelian Higgs model in Ref.~\cite{ae98}, we will assume, as there, that bubbles nucleate at points symmetrically placed on the $z$-axis at $\pm R$ at time $t=0$, that they expand from the point of nucleation with a speed approaching the speed of light, and that they collide at time  $t=t_c=R$.  They present their results in terms of the time after collision, which we will call $\delta t$, so that $t=tc+\delta t$ and $\tau=\sqrt{(\delta t+R)^2-\rho^2}$.

We write the overall normalization of the electromagnetic current by combining the factor containing the coupling parameters appearing explicitly in Eq.~(\ref{jfinal}) and the square of the normalization of the  $W^{\pm}$ fields appearing explicitly in Eqs.~(\ref{BCI} and \ref{BCII}). This normalization is thus
\beq
\label{currentnorm}
w^2\frac{gg'}{\sqrt{g^2+g'^2}} ~.
\eeq
Using the values of $g$ and $g'$ quoted below Eq.~(\ref{AZ}) and the value of $w$ specified in Eq.~(\ref{wnorm2}) of the Appendix, we find that Eq.(\ref{currentnorm}) becomes
\begin{eqnarray}
\label{currentnorm1}
w^2\frac{gg'}{\sqrt{g^2+g'^2}}&\approx& -38.5+1.36T_c~GeV \\ \nonumber
&=&2.32m_W ~,
\end{eqnarray}
taking $T_c= 166 GeV$ from Ref.~\cite{eikr}.

\begin{figure}
\centerline{\epsfig{file=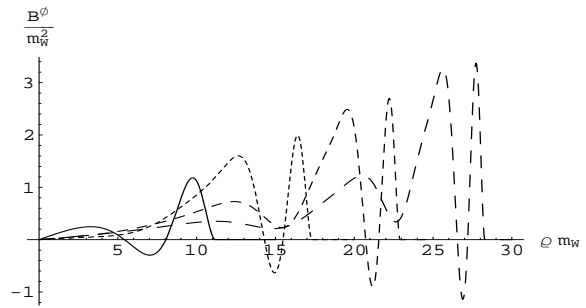,height=4cm,width=8.cm}}
\caption{Magnitude of the azimuthal magnetic field calculated in the theory of the present paper.  The field is shown as a function of distance $\rho$ from the axis of collision in the symmetry plane at time $\delta t=5,~10,~15$, and $20$.}
\label{ourb1newf}
\end{figure}

Figure.~\ref{ourb1newf} shows our calculated field at a sequence of times $\delta t$ after the time of collision assuming that the bubbles nucleated at a distance of $\pm 10$ from the origin ($R=10$).  The magnetic field shown in this figure is the value of the azimuthal field in the symmetry plane, which is the plane perpendicular to the axis of collision at the point where the bubbles first collide. Distance and time are expressed in units of $1/m_w$ and the magnetic field in units of $m_W^2$.  It is clear that the strength of the field is largest in the outermost region of the surface of the expanding overlap region and that this region narrows as the bubble expands.  The fact that the speed of expansion can be superluminal is not in contradiction with relativity, as discussed in Ref.~\cite{cst00}.

\begin{figure}
\centerline{\epsfig{file=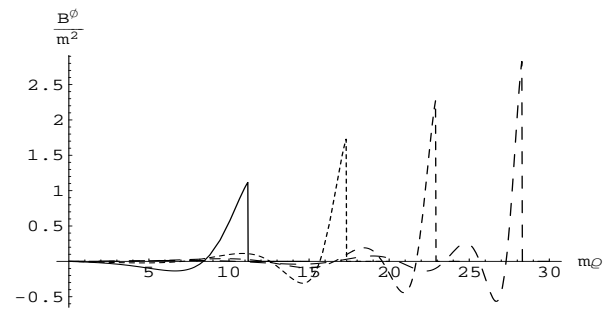,height=4cm,width=8.cm}}
\caption{Magnitude of the azimuthal magnetic field calculated in the  Abelian Higgs model. The field is shown as a function of distance $\rho$ from the axis of collision in the symmetry plane at time $\delta t=5,~10,~15$, and $20$.}
\label{aeb1newf}
\end{figure}

This is to be compared to the result in the  Abelian Higgs model, shown in Fig.~\ref{aeb1newf}.  A convenient expression for the magnetic field is given in Eq.~(29) of Ref.~\cite{cst00}. We have taken~\cite{ae98} $\theta_0=1$, the charge $e=1$, again expressing distance and time in units of inverse mass and the magnetic field in mass squared, where the mass in this case is that of the vector boson in that model.

There are several points to be made from the comparison of Fig.~\ref{ourb1newf} and Fig.~\ref{aeb1newf}.  The first point is that the magnitude of the magnetic field in the present model is about the same size as the one obtained in the  Abelian Higgs model.  This is the case even though the description of the physics is quite different in the two cases, with the source of the {\it em} current in the present model being charged gauge bosons that originate in the plasma.  Once these  $W^{\pm}$ pass through the bubble wall they (1) evolve in space and time according to our equations of motion and do so coherently with the Higgs field to which they couple; and, (2) populate the bubble preserving the number density to the extent allowed by overall energy-momentum conservation, as discussed in the appendix.

Secondly, it is clear that the magnetic field extends more deeply into the collision region in our theory, which makes the volume-averaged azimuthal field in the surface of the ring even larger than it is in the  Abelian Higgs model. This possible source of enhancement is however mitigated by the fact that the finite conductivity $\sigma$ damps the magnetic field in the interior of the ring.  The damping is effective beginning at depth $\delta\rho_{ring}$ from the surface of the ring,
\beq
\label{depth}
\delta\rho_{ring}=\frac{\sigma}{2m_W^2\rho_{ring}}~,
\eeq
where
\beq
\label{cond}
\sigma\approx 6.7T_c~.
\eeq
The ring expands with the time elapsed after the initial collision $\delta t$ as
\beq
\label{ring}
\rho_{ring}=\sqrt{2Rv\delta t}~,
\eeq
where $R$ is the radius of a bubble when the collision occurred and $v$ is the wall speed~\cite{kv95}.

Note also that the magnetic field in our theory drops to zero at the outer boundary of the expanding overlap region whereas it terminates suddenly in the  Abelian Higgs model.  This unphysical feature was discussed in Ref.~\cite{cst00} and was traced to the abrupt change in the boundary condition at the point of collision.  The same behavior occurs in our theory in the  $W^{\pm}$ fields, but its effect on the magnetic field is evidently mitigated when we evaluate the current with the charged  $W^{\pm}$ as its source.

\begin{figure}
\centerline{\epsfig{file=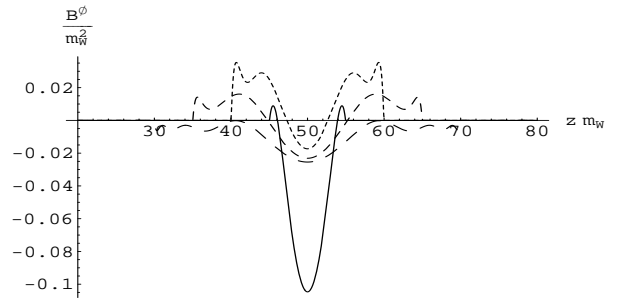,height=4cm,width=8.cm}}
\caption{Magnitude of the azimuthal magnetic field calculated in the theory of the present paper. The field is shown as a function of distance $z$ along the axis of collision at a distance $\rho=1$ from the axis of collision for times $\delta t=5,~10,~15$, and $20$.}
\label{ourbznewf}
\end{figure}

\begin{figure}
\centerline{\epsfig{file=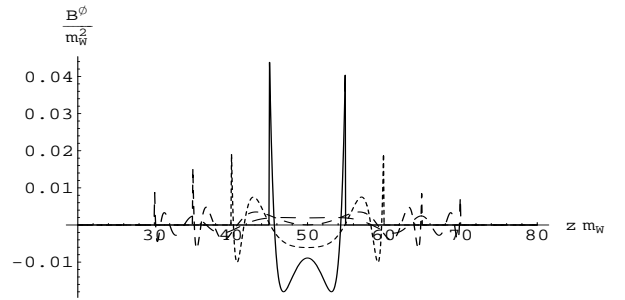,height=4cm,width=8.cm}}
\caption{Magnitude of the azimuthal magnetic field calculated in the  Abelian Higgs model. The field is shown as a function of distance $z$ along the axis of collision at a distance $\rho=1$ from the axis of collision for times $\delta t=5,~10,~15$, and $20$.}
\label{aebznewf}
\end{figure}

Next, we want to examine the magnetic field as we move along the $z$ direction at fixed radial distance $\rho$ from the collision axis.  By comparing the results shown in Fig.~\ref{ourbznewf} and Fig.~\ref{aebznewf} for our theory and that of the  Abelian Higgs model, respectively, we again see similar magnitudes of the magnetic fields.  The azimuthal magnetic field in the  Abelian Higgs model again reaches its peak at the surface of the intersection region, whereas in our theory is generally small there.  Thus, again we see that when we take the source of the electromagnetic current to be the charged  $W^{\pm}$ fields, the magnetic field created in collisions behaves in a smoother fashion than it does in the  Abelian Higgs model with step-function boundary conditions.  The tendency to peak at the edge of the intersection region makes strong longitudinal oscillations in $B$ in the  Abelian Higgs model, particularly when we look deep within the overlap region.  Because this region characterized by relatively small $\rho$ has had more time to evolve than the regions closer to the outer ring where $\rho$ is large, the conductivity has its most pronounced effect, strongly damping the fields as well as concentrating them into a single peak~\cite{ae98}.

The quantity we need to obtain from our calculation for determining the effectiveness of magnetic field creation during the electroweak phase transition seeding the galactic and extragalactic magnetic fields we see today is the azimuthal magnetic field strength $B(R)$ created at the surface of the ring that is formed by the colliding bubbles. In the  Abelian Higgs model, the value of $B(R)$ is given by Eq.~(24) of Ref.~\cite{ae98} as a function of $v_{wall}$, the radius of the bubbles $R$ at the time of collision $t=t_c$, and the radius $\rho_{ring}$ at the time of completion of the phase transition.  This result does not depend explicitly on the conductivity $\sigma$.  This is because the conductivity damps out the magnetic field in the interior of the bubble
but not at the outer surface of the ring where it is the largest.
However, the finite wall speed does damp the magnetic field at the surface as shown in Eq.~(24) of Ref.~\cite{ae98}.  Because this damping is a kinematic effect, it scales with the value of the magnetic field at the surface of the ring.  Hence, from knowledge of the value of the magnetic field in the ring, we can obtain the appropriate $B(R)$ for our theory by scaling it relative to the value of the magnetic field in the ring in the  Abelian Higgs model. It is clear from Fig.~\ref{ourb1newf}, Fig.~\ref{aeb1newf}, Fig.~\ref{ourbznewf}, and Fig.~\ref{aebznewf} that this scaling indicates that our seed fields will lead to about the same magnetic field as the estimate in Eq.~(30) of Ref.~\cite{ae98} for the galactic dynamo.

\section{Summary and Conclusions}
Methods suitable for exploring magnetic seed field creation during a first-order EW phase transition have been developed for the MSSM.  We obtain equations of motion from the EW Lagrangian, omitting fermions.
One of the most important consequences of these equations is that the fields of the non-abelian sector are the only charged ones that appear in the electromagnetic current, and therefore these fields (the charged gauge bosons $W^\pm$) are the sole source of the
magnetic seed fields.  This is in sharp contrast to the  Abelian Higgs model (\cite{kv95}, \cite{ae98}, \cite{cst00}), where the magnetic field arises from gradients in the phase of the Higgs field.

We studied numerically the magnetic fields produced in collisions of the bubbles of the first-order electroweak phase transition with our equations of motion and found, for gentle collisions (where fluctuations in the Higgs field are relatively small in bubble collisions), that the $W^{\pm}$ fields required for calculating the current may be obtained by solving {\it linear} equations.  Numerical solutions of the partial differential equations were obtained by applying simple boundary conditions that fix the fields at the time of bubble collision, similar to those employed in Refs.~\cite{kv95},~\cite{ae98}, and~\cite{cst00}. The magnetic seed fields we obtained with the charged  $W^{\pm}$ as their source are comparable to those estimated in the  Abelian Higgs model.

These results establish the electroweak phase transition in the MSSM as a promising source for production of seed fields for large-scale galactic and extra-galactic fields observed today. In our future work~\cite{stevens1} on magnetic field generation we will explore sensitivity to some of the more restrictive assumptions we make in this work.

\vspace{3mm}

\Large{{\bf Acknowledgements}}\\
\normalsize
MBJ and EMH acknowledge partial support of the DOE.  Special thanks go to to Los Alamos National Laboratory for its partial support of the PhD research of T. Stevens.
\vspace{2cm}

\appendix{{\bf Appendix: Normalization of Charged  $W^{\pm}$ Fields}}

The process by which thermal modes of the plasma make the transition into the bubble is a complicated many-body problem that we are unable to simulate.  In order to obtain numerical estimates, we introduce instead a model for determining the normalization of the charged  $W^{\pm}$ fields.  Our main assumption is that the number density of  $W^{\pm}$ quanta in bubble is the same as the number density of the quanta in the thermal plasma in the modes from which they originated.
The density of the  $W^{\pm}$ outside the bubble that are able to make the transition, $\rho_W$, is
\beq
\label{distr1}
\rho_W(T)=\int^{\bar k (T)}_0\frac{d^3k}{(2\pi)^3}\frac{1}{e^{k/T}-1} ~.
\eeq
where $\bar k$ is fixed by overall energy-momentum conservation. Energy conservation is an issue because in the plasma the  $W^{\pm}$ are massless (having energy $\hbar k$), whereas inside the bubble their energy is the same as their mass (see Eq.~(\ref{eomus12})). {\it Overall} energy-momentum conservation is a weaker constraint than insisting that energy and momentum be conserved in each collision, and is a valid constraint when multiple simultaneous collisions occur at the bubble wall, as we assume to happen at the high densities of interest here.  Application collision-by-collision will lead to a finite wall speed~\cite{dyne}.  The total energy of the  $W^{\pm}$ in a single bubble is given by their mass times the density in Eq.~(\ref{distr1}).

\begin{figure}
\centerline{\epsfig{file=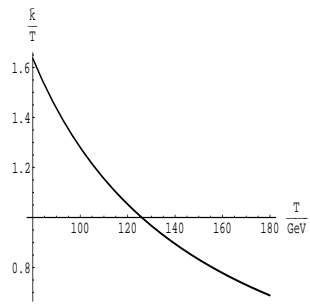,height=4cm,width=4.cm}}
\caption{Spectral limit}
\label{xbar}
\end{figure}

Overall energy conservation thus determines the portion of the spectrum $(k<\bar k)$ that populates the bubble as it grows,
\begin{eqnarray}
\label{kbar}
&m_W&2\int^{\bar k (T)}_0\frac{d^3k}{(2\pi)^3}\frac{1}{e^{k/T}-1} \\ \nonumber
&=&2\int^{\bar k (T)}_0k\frac{d^3k}{(2\pi)^3}\frac{1}{e^{k/T}-1}~.
\end{eqnarray}The factors of 2 arise from the fact that there are only two spin directions for massless vector bosons in the plasma. We show the solution ${\bar k}(T)$ of Eq.~(\ref{kbar}) in Fig.~\ref{xbar}. In the region of temperature shown, $\bar k\approx 3m_W/2$.
The density $\rho_W$ of  $W^{\pm}$ inside the bubble is then

\beq
\label{distr}
\rho_W(T)=\frac{T^3}{\pi^2}\int^{\bar k (T)/T}_0 \frac{ x^2dx }{e^x-1}~.
\eeq

The number of quanta filling the bubble clearly grows as the bubble expands and displaces a larger volume of the plasma. Thus, inside the bubble the number of  $W^{\pm}$ boson pairs $N$ is $N=\rho_W(T)\Omega$, where $\Omega=4\pi/3(R_0^2+t^2)^{3/2}$ is the volume of the bubble.

For a single  $W^{\pm}$, assuming relativistic normalization of wave functions,
\beq
\label{wnorm1}
w=1/\sqrt{2m_w\Omega}~,
\eeq
where $\Omega$ is the volume of the bubble.  Therefore, for $N$  $W^{\pm}$ bosons in a bubble,
\beq
\label{wnorm2}
w=\sqrt{N}/\sqrt{2m_w\Omega}=\sqrt{\frac{\rho_W(T)}{2m_W}}~.
\eeq
For $\bar k$ given in Fig.~\ref{xbar},  $w$ is nearly linear in temperature and is given over the same temperature range as
\beq
\label{wt}
w(T)^2=-127.+4.49T
\eeq
in units of GeV.  It is unlikely that more $W$ from the plasma than allowed by energy-momentum conservation would drop into the coherent mode since the excess energy would lead to heating that would impede the phase transition.



\end{document}